\documentclass[reprint,prd,showpacs]{revtex4-1}
\usepackage{amsmath}
\usepackage{mathtools}
\usepackage{upgreek}
\usepackage{textgreek}
\usepackage{graphicx}
\usepackage{verbatim}
\usepackage[final]{changes}
\usepackage{microtype}
\usepackage{float}
\PassOptionsToPackage{hyphens}{url}\usepackage[pdftex,draft=false,
			pdfauthor={Antonio Mondragon},
			pdfcreator={pdflatex},
			pdftitle={Kepler's Orbits and Special Relativity in Introductory Classical Mechanics},
			pdfproducer={Latex with hyperref},
			pdfkeywords={
			43,14.3,7.16,
			angular momentum,
			approximate,
			arc seconds per century,
			central mass,
			circular,
			centrifugal barrier,
			classical mechanics,
			change of variable,
			conservation,
			coordinate transformation,
			corrections,
			correspondence principle,
			eccentric, 
			eccentricity,
			effective potential,
			ellipse,
			elliptical,
			general relativity,
			gravity, 
			gravitation,
			gravitational potential,
			increased eccentricity,
			kepler,
			Keplerian,
			keplerian limit,
			Lagrangian,
			Mercury,
			near-Keplerian,
			Newtonian,
			orbit,
			orbital period,
			perihelia, 
			perihelion, 
			perihelic,
			perturbation, 
			perturbative, 
			perturbatively,
			physics majors,
			planets,
			precession,
			precession of perihelion,
			precession of perihela,
			radial,
			reduced radius,
			relativistic corrections,
			relativistic capture,
			relativistic mass,
			relativity,
			small velocity,
			solar system,
			special relativity, 
			spacetime,
			speed of light,
			toy model,
			unstable,
			velocity-dependent,
			Einstein,
			Lagrange,
			Lemmon,
			Mondragon,
			Newton,
			Schwarzschild}]{hyperref}
			\usepackage[all]{hypcap}
\hypersetup{linktoc=all,
			bookmarksopen=true,
            bookmarksnumbered=true,
            breaklinks=true,
            colorlinks=true,
			linkcolor=black,
			filecolor=black,
			urlcolor=black,
			citecolor=black}
\def\rmd{\mathrm{d}}
\def\bs{\boldsymbol}
\def\gammadot{\dot{\gamma\phantom{\,}}\!}
\def\tildegamma{\bar{\gamma\phantom{\,}}\!}

\def\baselinestretch{1.1075}
\makeatletter
\g@addto@macro{\UrlBreaks}{\UrlOrds}
\makeatother
\begin{document}

\title{Kepler's Orbits and Special Relativity \\ in \\ Introductory Classical Mechanics}
\date{\today}
\author{Tyler J. Lemmon}
\altaffiliation{Colorado College}
\author{Antonio R. Mondragon}
\altaffiliation{Colorado College}
\email{antonio.mondragon@gmail.com}

\begin{abstract}
Kepler's orbits with corrections due to Special Relativity are explored using the Lagrangian formalism.  A very simple model includes only relativistic kinetic energy by defining a Lagrangian that is consistent with both the relativistic momentum of Special Relativity and Newtonian gravity.  The corresponding equations of motion are solved in a Keplerian limit, resulting in an approximate relativistic orbit equation that has the same form as that derived from General Relativity in the same limit and clearly describes three characteristics of relativistic Keplerian orbits:  precession of perihelion; reduced radius of circular orbit; and increased eccentricity.  The prediction for the rate of precession of perihelion is in agreement with established calculations using only Special Relativity.  All three characteristics are qualitatively correct, though suppressed when compared to more accurate general-relativistic calculations.  This model is improved upon by including relativistic gravitational potential energy.  The resulting approximate relativistic orbit equation has the same form and symmetry as that derived using the very simple model, and more accurately describes characteristics of relativistic orbits.  For example, the prediction for the rate of precession of perihelion of Mercury is one-third that derived from General Relativity.  These Lagrangian formulations of the special-relativistic Kepler problem are equivalent to the familiar vector calculus formulations.  In this Keplerian limit, these models are supposed to be physical based on the likeness of the equations of motion to those derived using General Relativity.  The resulting approximate relativistic orbit equations are useful for a qualitative understanding of general-relativistic corrections to Keplerian orbits.  The derivation of this orbit equation is approachable by undergraduate physics majors and nonspecialists whom have not had a course dedicated to relativity.
\end{abstract}
\maketitle
\thispagestyle{empty}

%%%%%%%%%%%%%%%%%%%%%%%%%
\section{Introduction} \label{sec_intro}
%%%%%%%%%%%%%%%%%%%%%%%%%

The relativistic contribution to the rate of precession of perihelion of Mercury is calculated accurately using General Relativity \cite{einstein0,einstein1,einstein2,einstein3,schwarzschild,droste}.  However, the problem is commonly discussed in undergraduate and graduate classical mechanics textbooks, without introduction of an entirely new, metric theory of gravity.  One approach \hbox{\cite{goldstein2,jose,peters,*phipps1,*phipps2,jia}} is to define a Lagrangian that is consistent with both the momentum-velocity relation of Special Relativity and Newtonian gravity.  The resulting equations of motion are solved perturbatively, and an approximate rate of precession of perihelion of Mercury is extracted.  This approach is satisfying in that a familiar element of Special Relativity---relativistic momentum---produces a small modification to a familiar problem---Kepler's orbits---and results in a characteristic of general-relativistic orbits---precession of perihelion.  On the other hand, one must be content with an approximate rate of precession that is one-sixth the correct value.  Another approach \cite{goldstein,TM1,barger,fowles,hand} is that of a history lesson and mathematical exercise.  A modification to Newtonian gravity is postulated, resulting in an equation of motion that is the same as that derived from General Relativity.  The equation of motion is solved perturbatively, and the correct rate of precession of perihelion of Mercury is extracted.  This method is satisfying in that the modification to Newtonian gravity results in the observed value for the relativistic contribution to perihelic precession.  On the other hand, one must be content with a mathematical exercise, rather than an understanding of the metric theory of gravity from which the modification of Newtonian gravity is derived.  Both approaches provide an opportunity for students of introductory classical mechanics to learn that relativity is responsible for a small contribution to perihelic precession and to calculate that contribution.

A review of the approach using only Special Relativity and an alternative solution of the equations of motion in a Keplerian limit are presented---resulting in an approximate relativistic orbit equation.  This orbit equation has the same form as that derived using General Relativity and clearly describes three relativistic corrections to Keplerian orbits:  precession of perihelion, reduced radius of circular orbit, and increased eccentricity.  The approximate rate of perihelic precession is in agreement with established calculations using only Special Relativity.  The method of solution makes use of a simple change of variables and the correspondence principle, rather than standard perturbative techniques, and is approachable by undergraduate physics majors.

Two models are considered.  A very simple model (Secs.\,\ref{sec_lagrange}~and~\ref{Sec_KeplerianLimit}) consists of relativistic kinetic energy and unmodified Newtonian gravity.  An improved model (Sec.\,\ref{sec_vector}) includes both relativistic kinetic energy and relativistic modification to Newtonian gravity.  A special-relativistic force is approximated in a Keplerian limit, resulting in a conservative approximate relativistic force, from which a relativistic potential energy is derived---thereby enabling the usage of the Lagrangian formalism.  For both models, the Lagrangian formalism is demonstrated to be equivalent to the vector calculus formalism.  The keplerian limit and validity of the approximate relativistic orbit equations is discussed thoroughly in Sec.\,\ref{sec_discussion}.  These models are supposed to be physical based on the likeness of the equations of motion to those derived using General Relativity (Sec.\,\ref{sec_structure}).  The physical relevance of these models is emphasized in Sec.\,\ref{sec_characteristics}.  A method of construction of similar models using more general modifications to Newtonian gravity is discussed, including a more broadly-defined Keplerian limit that is useful in the derivation of orbit equations for such models (App.\,\ref{app_math_prob} and Sec.\,\ref{sec_discuss_toy}).

%%%%%%%%%%%%%%%%%%%%%%%%%
\section{Relativistic Kinetic Energy} \label{sec_lagrange}
%%%%%%%%%%%%%%%%%%%%%%%%%

Conceptually, the simplest relativistic modification to Kepler's orbits is to define a Lagrangian with a kinetic energy term that is consistent with both the momentum-velocity relation of Special Relativity and the Newtonian gravitational potential energy \cite{goldstein2,jose,peters,*phipps1,*phipps2,TM2,*barger2,potgieter,DesEri,HuangLin,SonMas,LemMon2};
\begin{equation}
L = -mc^2 \gamma^{-1} + \frac{GMm}{r}, \label{eq_lagrangian}
\end{equation}
where $\gamma^{-1} \equiv \sqrt{1 - v^2/c^2}$, and $v^2 = \dot{r}^2 + r^2 \dot{\theta}^2$.  ($G$ is Newton's universal gravitational constant, $M$ is the mass of the sun, and $c$ is the speed of light in vacuum.)  The equations of motion follow from Lagrange's equations
\begin{equation}
\frac{\rmd}{\rmd t} \frac{\partial L}{\partial\dot{q}_i} - \frac{\partial L}{\partial q_i} = 0,
\end{equation}
where $\dot{q_i} \equiv \rmd q_i/\rmd t$ for each of $\{ q_i \} = \{ \theta,r \}$.  The results are
\begin{gather}
\dfrac{\rmd}{\rmd t} ( \gamma r^2 \dot{\theta} ) = 0, \label{eq_der_ang_mom} \\
\text{and} \nonumber \\
\gamma \ddot{r} + \gammadot \dot{r} + \frac{GM}{r^2} - \gamma r \dot{\theta}^2 = 0. \label{eq_eom}
\end{gather}
Using Eq.\,(\ref{eq_der_ang_mom}), a relativistic analogue to the Newtonian equation for conservation of angular momentum per unit mass is defined
\begin{equation}
\ell \equiv \gamma r^2 \dot{\theta} = \text{constant}. \label{eq_ang_mom}
\end{equation}
This is used to eliminate the explicit occurrence of $\dot{\theta}$ in the equation of motion Eq.\,(\ref{eq_eom})
\begin{equation}
\gamma r \dot{\theta}^2 = \frac{\ell^2}{\gamma r^3}. \label{eq_part1}
\end{equation}
Time is eliminated by successive applications of the chain rule, together with conserved angular momentum \cite{goldstein4,*jose2,*TM4,*fowles2,hamill};
\begin{equation}
\dot{r} = - \frac{\ell}{\gamma} \frac{\mathrm{d}}{\mathrm{d} \theta} \frac{1}{r}, \label{eq_r_dot_mass}
\end{equation}
and, therefore,
\begin{equation}
\gamma \ddot{r} = - \gammadot \dot{r} - \frac{\ell^2}{\gamma r^2} \frac{\mathrm{d^2}}{\mathrm{d} \theta^2} \frac{1}{r}. \label{eq_part2}
\end{equation}
Substituting Eqs.\,(\ref{eq_part1})~and~(\ref{eq_part2}) into the equation of motion Eq.\,(\ref{eq_eom}) results in
\begin{equation}
\ell^2\frac{\mathrm{d}^2}{\mathrm{d} \theta^2} \frac{1}{r} - \gamma GM + \frac{\ell^2}{r} = 0. \label{eq_EOM2_mass}
\end{equation}
Anticipate a solution of Eq.\,(\ref{eq_EOM2_mass}) that is near Keplerian and introduce the radius of a circular orbit for a nonrelativistic particle with the same angular momentum, $r_\text{c} \equiv \ell^2/GM $.  The result is
\begin{equation}
\frac{\mathrm{d}^2}{\mathrm{d} \theta^2} \frac{r_\text{c}}{r} + \frac{r_\text{c}}{r} = 1 + \lambda, \label{eq_SR_rel_mass}
\end{equation}
where $\lambda\equiv \gamma - 1$ is a velocity-dependent correction to Newtonian orbits due to Special Relativity.  The conic sections of Newtonian mechanics \cite{goldstein3,*jose3,*TM3,*fowles3,*hamill2,smiths} are recovered by setting $\lambda = 0$ $(c \rightarrow \infty)$
\begin{equation}
\frac{\mathrm{d}^2}{\mathrm{d} \theta^2} \frac{r_\text{c}}{r} + \frac{r_\text{c}}{r} = 1, \label{eq_Newton1}
\end{equation}
resulting in the well-known orbit equation
\begin{equation}
\frac{r_\text{c}}{r} = 1 + e\cos{\theta}, \label{eq_Newton2}
\end{equation}
where $e$ is the eccentricity.  Kepler's orbits are described by $0 < e < 1$.

%%%%%%%%%%%%%%%%%%%%%%%%%
\section{Keplerian Limit and Orbit Equation}\label{Sec_KeplerianLimit}
%%%%%%%%%%%%%%%%%%%%%%%%%

The planets of our solar system are described by near-circular orbits $(e \ll 1)$ and require only small relativistic corrections $(v/c \ll 1)$.  Mercury has the largest eccentricity $(e \approx 0.2)$, and the next largest is that of Mars $(e \approx 0.09)$.  Therefore, $\lambda$ [defined after Eq.\,(\ref{eq_SR_rel_mass})] is taken to be a small relativistic correction to near-circular orbits of Newtonian mechanics---Keplerian orbits.  This correction is approximated using the first-order series $\gamma \approx 1 + \tfrac{1}{2} (v/c)^2$, and neglecting the radial component of velocity $v \approx r \dot{\theta}$ 
\begin{equation}
\lambda \approx \tfrac{1}{2} (r\dot{\theta}/c)^2. \label{eq_approx_lambda2}
\end{equation}
See Sec.\,\ref{sec_discussion} for a thorough discussion of this approximation.  Using angular momentum Eq.\,(\ref{eq_ang_mom}) to eliminate $\dot{\theta}$ results in $\lambda \approx \tfrac{1}{2}  (\ell/rc)^2 (1 + \lambda)^{-2}$, or
\begin{equation}
\lambda \approx \tfrac{1}{2}  (\ell/rc)^2.
\end{equation}
The equation of motion Eq.\,(\ref{eq_SR_rel_mass}) is now expressed approximately as
\begin{equation}
\frac{\mathrm{d}^2}{\mathrm{d} \theta^2} \frac{r_\text{c}}{r} + \frac{r_\text{c}}{r} \approx 1 + \tfrac{1}{2}  \epsilon \left( \dfrac{r_\text{c}}{r} \right)^{\!2}. \label{eq_SR9}
\end{equation}
where $\epsilon \equiv (GM/\ell c)^2$.  The conic sections of Newtonian mechanics, Eqs.\,(\ref{eq_Newton1})~and~(\ref{eq_Newton2}), are now recovered by setting $\epsilon = 0$ $(c \rightarrow \infty)$.  The solution of Eq.\,(\ref{eq_SR9}) for $\epsilon \ne 0$ approximately describes Keplerian orbits with small corrections due to Special Relativity.

%%%%%%%%%%%%%%%%%%%%

If $\epsilon$ is taken to be a small relativistic correction to Keplerian orbits, it is convenient to make the change of variable $1/s \equiv {r_\text{c}}/{r} - 1 \ll 1$.  The last term on the right-hand-side of Eq.\,(\ref{eq_SR9}) is then approximated as \hbox{$(r_\text{c}/r )^2 \approx 1 + 2/s$,} resulting in a linear differential equation for $1/s(\theta)$
\begin{equation}
\frac{\rmd^2}{\rmd \theta^2} \frac{2}{\epsilon s} + \frac{ 2( 1 - \epsilon) }{\epsilon s} \approx 1. \label{eq_linearized2}
\end{equation}
The additional change of variable $\alpha \equiv \theta \sqrt{1 - \epsilon}$ results in the familiar form
\begin{equation}
\frac{\mathrm{d}^2}{\mathrm{d} {\alpha}^2} \frac{s_\text{c}}{s} + \frac{s_\text{c}}{s} \approx 1, \label{eq_SR3}
\end{equation}
where $s_\text{c} \equiv 2(1 - \epsilon)/\epsilon$.  The solution is similar to that of Eq.\,(\ref{eq_Newton1})
\begin{equation}
\frac{s_\text{c}}{s} \approx 1   + A\cos{\alpha},\label{eq_SR4}
\end{equation}
where $A$ is an arbitrary constant of integration.  In terms of the original coordinates
\begin{equation}
\frac{\bar{r}_\text{c}}{r} \approx 1 + \bar{e} \cos{\bar{\kappa}\theta}, \label{eq_S_eoo}
\end{equation}
where
\begin{align}
\bar{r}_\text{c} &\equiv r_\text{c}\frac{1 - \epsilon}{1 - \tfrac{1}{2} \epsilon} \label{eq_S_coeff_r0}\\
\bar{e} &\equiv \frac{\tfrac{1}{2} \epsilon A}{1 - \tfrac{1}{2}  \epsilon} \label{eq_S_coeff_e0}\\
\bar{\kappa} &\equiv (1 - \epsilon)^{\frac{1}{2}}. \label{eq_S_coeff_phi0}
\end{align}
According to the correspondence principle, Kepler's orbits [Eq.\,(\ref{eq_Newton2}) with $0 < e < 1$] must be recovered in the limit $\epsilon\rightarrow 0$ $(c \rightarrow \infty)$, so that $\tfrac{1}{2} \epsilon A \equiv e$ is the eccentricity of Newtonian mechanics. To first order in $\epsilon$, Eqs.\,(\ref{eq_S_coeff_r0})--(\ref{eq_S_coeff_phi0}) are
\begin{align}
\bar{r}_\text{c} &\approx r_\text{c} (1 - \tfrac{1}{2}\epsilon) \label{eq_S_coeff_r}\\
\bar{e} &\approx e (1 + \tfrac{1}{2}\epsilon) \label{eq_S_coeff_e} \\
\bar{\kappa} &\approx 1 - \tfrac{1}{2}\epsilon, \label{eq_S_coeff_phi}
\end{align}
so that relativistic orbits in this limit are described concisely by
\begin{equation}
\frac{r_\text{c}(1 - \tfrac{1}{2}\epsilon)}{r} \approx 1 + e(1 + \tfrac{1}{2}\epsilon)\cos{(1 - \tfrac{1}{2}\epsilon)\theta}. \label{eq_class_rel}
\end{equation}
When compared to Kepler's orbits [Eq.\,(\ref{eq_Newton2}) with \hbox{$0 < e < 1$],} this orbit equation clearly displays three characteristics of near-Keplerian orbits: precession of perihelion; reduced radius of circular orbit; and increased eccentricity.  This approximate orbit equation has the same form as that derived from General Relativity in this limit \cite{LemMon}
\begin{equation}
\frac{r_\text{c}(1 - 3 \epsilon)}{r} \approx 1 + e(1 + 3 \epsilon)\cos{(1 - 3 \epsilon)\theta}. \label{eq_gen_rel}
\end{equation}

The equations of motion, Eqs.\,(\ref{eq_der_ang_mom})~and~(\ref{eq_eom}), are identical to those derived using the simple force equation (Newton's 3$^\text{rd}$ Law) \hbox{$\dot{\textbf{p}} = - GMm\hat{\textbf{r}}/r^2$,} where \hbox{$\textbf{p} = p_r \hat{\textbf{r}} + p_\theta \hat{\bs{\uptheta}}$,} and \hbox{$\{ p_r,p_\theta \} = \{ \gamma m \dot{r}, \gamma m r \dot{\theta} \}$,} verifying that the unfamiliar relativistic kinetic energy term in the Lagrangian, \hbox{$\tilde{T} \equiv -mc^2 \gamma^{-1}$} in Eq.\,(\ref{eq_lagrangian}), is consistent with the familiar definition of relativistic momentum \hbox{$\textbf{p} = \gamma m \textbf{v}$.}  For example, with \mbox{$\dot{\textbf{p}} = \dot{p}_r \hat{\textbf{r}} + \dot{p}_\theta \hat{\bs{\uptheta}}$,} and using \hbox{$\{ \rmd \hat{\textbf{r}}/\rmd t, \rmd \hat{\bs{\uptheta}}/\rmd t \} = \{ \hat{\bs{\uptheta}} \dot{\theta}, -\hat{\textbf{r}} \dot{\theta} \}$}, the conserved relativistic angular momentum Eq.\,(\ref{eq_der_ang_mom}) is verified
\begin{equation}
\dot{p}_\theta/m = \frac{1}{r} \dfrac{\rmd}{\rmd t} ( \gamma r^2 \dot{\theta} ) = 0.
\end{equation}

%%%%%%%%%%%%%%%%%%%%%%%%%
\section{Relativistic Gravitational Potential Energy} \label{sec_vector}
%%%%%%%%%%%%%%%%%%%%%%%%%

The effect of using the special-relativistic $\gamma$ factor to define both relativistic kinetic energy and relativistic gravitational potential energy is explored \cite{SinghPatra,bunchaft,hidalgo,phipps3,kurucz,vfg,abci}.  A relativistic gravitational potential energy is derived from a conservative approximate gravitational force.  In this Keplerian limit, a Lagrangian formulation of this problem is demonstrated to be equivalent to the vector calculus formulation.

%%%%%%%%%%%%%%%%%%%%%%%%%
\subsection{Vector Calculus Formalism} \label{sec_vector_form}
%%%%%%%%%%%%%%%%%%%%%%%%%

The central-mass problem including both relativistic kinetic energy and relativistic gravitational force is described by the simple force equation
\begin{equation}
\dot{\textbf{p}} = -\gamma \frac{GMm}{r^2}\hat{\textbf{r}},
\end{equation}
where $\textbf{p} = \gamma m \textbf{v}$.  The equations of motion are derived as described in the final paragraph in Sec.\,\ref{Sec_KeplerianLimit} and are identical to Eqs.\,(\ref{eq_der_ang_mom})~and~(\ref{eq_eom}), except that the gravitational force is multiplied by the relativistic $\gamma$ factor
\begin{gather}
\ell \equiv \gamma r^2 \dot{\theta} = \text{constant}, \label{eq_ang_mom2} \\
\text{and} \nonumber \\
\gamma \ddot{r} + \gammadot \dot{r} + \gamma \frac{GM}{r^2} - \gamma r \dot{\theta}^2. \label{eq_eom2}
\end{gather}
An approximate relativistic orbit equation is derived as described in Secs.\,\ref{sec_lagrange}~and~\ref{Sec_KeplerianLimit}
\begin{equation}
\frac{\mathrm{d}^2}{\mathrm{d} \theta^2} \frac{r_\text{c}}{r} + \frac{r_\text{c}}{r} = 1 + \bar{\lambda}, \label{eq_SR_rel_mass2}
\end{equation}
where $\bar{\lambda} \equiv \gamma^2 - 1 \approx (r \dot{\theta}/c)^2$.  Notice that $\bar{\lambda} = 2\lambda$, where $\lambda \equiv \gamma -1 \approx \tfrac{1}{2} (r \dot{\theta}/c)^2$ is defined after Eq.\,(\ref{eq_SR_rel_mass}) and in Eq.\,(\ref{eq_approx_lambda2}).  Therefore, the orbit equation is found using the simple replacement $\epsilon \rightarrow 2\epsilon$ in Eq.\,(\ref{eq_class_rel})
\begin{equation}
\frac{r_\text{c}(1 - \epsilon)}{r} \approx 1 + e(1 + \epsilon)\cos{(1 - \epsilon)\theta}. \label{eq_class_rel_mass}
\end{equation}
This approximate relativistic orbit equation has the same form as that derived in Secs.\,\ref{sec_lagrange}~and~\ref{Sec_KeplerianLimit} using only relativistic kinetic energy Eq.\,(\ref{eq_class_rel}) and is more accurate---when compared to that derived from General Relativity in this same limit Eq.\,(\ref{eq_gen_rel}).

%%%%%%%%%%%%%%%%%%%%%%%%%
\subsection{Lagrangian Formalism} \label{sec_lagrange_form}
%%%%%%%%%%%%%%%%%%%%%%%%%

An equivalent Lagrangian formulation of this problem is accomplished by defining a conservative approximate relativistic gravitational force, from which a corresponding potential energy is derived.  A relativistic gravitational force is defined as the Newtonian force due to gravity with the replacement $m \rightarrow \gamma m$
\begin{equation}
\tilde{F}_\text{g} \equiv -\gamma \frac{GMm}{r^2} = -\frac{GMm}{r^2}(1 + \lambda), \label{eq_rel_grav_force}
\end{equation}
where $\lambda \equiv \gamma - 1$ is a small correction to Newtonian gravity due to Special Relativity.  Also, relativistic angular momentum (per unit mass) is defined to be that of Newtonian mechanics with the replacement \hbox{$m \rightarrow \gamma m$,} \hbox{$\ell \equiv \gamma r^2 \dot{\theta}$.}  Anticipating near-circular orbits, $\lambda$ is approximated using the first-order series $\gamma \approx 1 + \tfrac{1}{2} (v/c)^2$, neglecting the radial component of velocity $v \approx r \dot{\theta}$, and using angular momentum to eliminate $\dot{\theta}$, as described in the first paragraph in Sec.\,\ref{Sec_KeplerianLimit}
\begin{equation}
\lambda \approx \tfrac{1}{2} \epsilon \left( \frac{r_\text{c}}{r} \right)^2.
\end{equation}
The result is a conservative approximate relativistic gravitational force Eq.\,(\ref{eq_rel_grav_force}) that \replaced{describes}{includes} a \replaced{small correction to Kepler's orbits due to Special Relativity}{term proportional to $c^{-2} r^{-4}$}
\begin{equation}
\tilde{F}_\text{g} \approx -\frac{GMm}{r^2} \left[ 1 + \tfrac{1}{2} \epsilon \left( \frac{r_\text{c}}{r} \right)^2 \right]. \label{eq_rel_grav_force2}
\end{equation}
This force is integrated to derive an approximate relativistic gravitational potential energy
\begin{equation}
\tilde{U}(r) = - \frac{GMm}{r} \left[ 1 + \tfrac{1}{6} \epsilon \left( \frac{r_\text{c}}{r} \right)^2 \right].
\end{equation}
The Lagrangian
\begin{equation}
L = -m c^2 \gamma^{-1} - \tilde{U}(r)
\end{equation}
results in the equations of motion
\begin{gather}
\dfrac{\rmd}{\rmd t} ( \gamma r^2 \dot{\theta} ) = 0, \label{eq_der_ang_mom3} \\
\text{and} \nonumber \\
\gamma \ddot{r} + \gammadot \dot{r} + \frac{GM}{r^2} \left[ 1 + \tfrac{1}{2} \epsilon \left( \frac{r_\text{c}}{r} \right)^2 \right] - \gamma r \dot{\theta}^2. \label{eq_eom3}
\end{gather}
The first equation Eq.\,(\ref{eq_der_ang_mom3}) verifies the definition of relativistic angular momentum, and the second equation Eq.\,(\ref{eq_eom3}) includes a small relativistic correction to \replaced{Kepler's orbits}{Newtonian gravity} due to Special Relativity.  Compare these equations of motion to those derived using only relativistic kinetic energy in Sec.\,\ref{sec_lagrange}, Eqs.\,(\ref{eq_der_ang_mom})~and~(\ref{eq_eom}).  A relativistic orbit equation is derived as described in Secs.\,\ref{sec_lagrange}~and~\ref{Sec_KeplerianLimit}
\begin{equation}
\frac{\mathrm{d}^2}{\mathrm{d} \theta^2} \frac{r_\text{c}}{r} + \frac{r_\text{c}}{r} = 1 + \bar{\lambda}, \label{eq_SR_rel_mass3}
\end{equation}
where $\bar{\lambda} \equiv \gamma [ 1 + \tfrac{1}{2} \epsilon (r_\text{c}/r)^2 ] - 1$ is a small correction to Keplerian orbits due to Special Relativity.  Using the first-order series $\gamma \approx 1 + \tfrac{1}{2} (v/c)^2$, neglecting the radial component of velocity $v \approx r \dot{\theta}$, and keeping terms first order in $\epsilon$ results in $\bar{\lambda} \approx 2\lambda$, where $\lambda \equiv \gamma -1 \approx \tfrac{1}{2} (r \dot{\theta}/c)^2$ is defined after Eq.\,(\ref{eq_SR_rel_mass}) and in Eq.\,(\ref{eq_approx_lambda2}).  Therefore, the orbit equation is found using the simple replacement $\epsilon \rightarrow 2\epsilon$ in Eq.\,(\ref{eq_class_rel}), thereby reproducing the orbit equation Eq.\,(\ref{eq_class_rel_mass}) derived using the vector calculus formalism in Sec.\,\ref{sec_vector_form}
\begin{equation}
\frac{r_\text{c}(1 - \epsilon)}{r} \approx 1 + e(1 + \epsilon)\cos{(1 - \epsilon)\theta}. \label{eq_class_rel_mass2}
\end{equation}
Characteristics of near-Keplerian orbits are most easily understood by comparing the approximate relativistic orbit equations to that derived from the nonrelativistic Kepler problem Eq.\,(\ref{eq_Newton2}).  For this purpose, it is convenient to express the orbit equation as
\begin{equation}
\frac{\bar{r}_\text{c}}{r} \approx 1 + \bar{e} \cos{\bar{\kappa}\theta}, \label{eq_S_eoo2}
\end{equation}
where, using the simple replacement $\epsilon \rightarrow 2\epsilon$ in Eqs.\,(\ref{eq_S_coeff_r0})--(\ref{eq_S_coeff_phi}),
\begin{alignat}{2}
\bar{r}_\text{c} &\equiv r_\text{c}\frac{1 - 2\epsilon}{1 - \epsilon} &&\approx r_\text{c} (1 - \epsilon) \label{eq_S_coeff_r_mass} \\
\bar{e} &\equiv e \frac{1}{1 - \epsilon} 	&&\approx e (1 + \epsilon) \label{eq_S_coeff_e_mass} \\
\bar{\kappa} &\equiv (1 - 2\epsilon)^{\frac{1}{2}} 	&&\approx 1 - \epsilon. \label{eq_S_coeff_phi_mass}
\end{alignat}

%%%%%%%%%%%%%%%%%%%%%%%%%
\section{Characteristics of Near-Keplerian Orbits} \label{sec_characteristics}
%%%%%%%%%%%%%%%%%%%%%%%%%

The approximate relativistic orbit equation Eq.\,(\ref{eq_class_rel_mass2}) predicts a shift in perihelion through an angle
\begin{equation}
\Delta\theta \equiv 2\pi (\bar{\kappa}^{-1} - 1) \approx 2\pi \epsilon \label{eq_S_def_precess}
\end{equation}
\begin{figure}%[H]
\centering\includegraphics{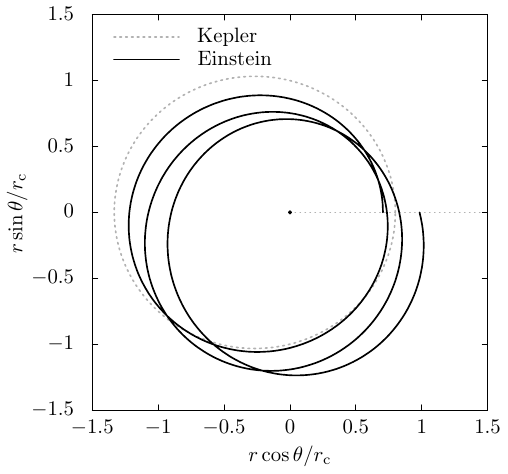}
\caption{\label{fig1} A relativistic orbit in a Keplerian limit (solid) Eq.\,(\ref{eq_class_rel_mass2}) is compared to a Keplerian orbit (dotted) Eq.\,(\ref{eq_Newton2}) with the same angular momentum.  Precession of perihelion is one characteristic of relativistic orbits and is illustrated here for \hbox{$0 \leq \theta \leq 6\pi$.}  Precession of perihelion is also predicted by General Relativity Eq.\,(\ref{eq_gen_rel}) in greater magnitude.  This characteristic of relativistic orbits is exaggerated by both the choice of eccentricity $(e=0.25)$ and relativistic correction parameter $(\epsilon=0.1)$ for purposes of illustration. Precession is present for smaller (non-zero) reasonably chosen values of $e$ and $\epsilon$ as well.  (The same value of $e$ is chosen for both orbits.)}
\end{figure}%
\noindent per revolution.  This prediction is twice that derived using the standard approach \hbox{\cite{goldstein2,jose,peters,*phipps1,*phipps2,frisch}} to incorporating Special Relativity into the Kepler problem, and is compared to observations assuming that relativistic and Keplerian angular momenta are approximately equal. For a Keplerian orbit \cite{goldstein3,*jose3,*TM3,*fowles3,*hamill2,smiths} $\ell^2 = GMa(1-e^2)$, where \hbox{$G = 6.670\times 10^{-11}\,\text{m}^3/\text{kg}\!\cdot\!\text{s}^2$,} \hbox{$M = 1.989\times 10^{30}$\,kg} is the mass of the Sun, and $a$ and $e$ are the semimajor axis and eccentricity of the orbit, respectively. Therefore, the relativistic correction defined after Eq.\,(\ref{eq_SR9}),
\begin{equation}
\epsilon \approx \frac{GM}{c^2 a (1 - e^2)},
\end{equation}
is largest for planets closest to the Sun and for planets with very eccentric orbits. For Mercury \cite{carroll,PleKra} \hbox{$a = 5.79 \times 10^{10}$\,m} and $e = 0.2056$, so that $\epsilon \approx 2.66\times 10^{-8}$. (The speed of light is taken to be \hbox{$c^2 = 8.987554\times 10^{16}\,\text{m}^2/\text{s}^2$.)} According to Eq.\,(\ref{eq_S_def_precess}), Mercury precesses through an angle
\begin{equation}
\Delta\theta \approx \frac{2\pi GM}{c^2 a (1 - e^2)} = 1.67 \times 10^{-7}\,\text{rad}
\end{equation}
per revolution. This angle is very small and is usually expressed cumulatively in arc seconds per century. The orbital period of Mercury is 0.24085 terrestrial years, so that
\begin{align}
\Delta\Theta &\equiv \dfrac{100\,\text{yr}}{0.24085\,\text{yr}} \times \dfrac{360 \times 60 \times 60}{2\pi} \times \Delta\theta \label{eq_prec_num_def} \\
&\approx 14.3\,\text{arcsec/century}. \label{eq_prec_num}
\end{align}
Precession, as predicted by Special Relativity is illustrated in Fig.~\ref{fig1}.
\begin{figure}%[H]
\centering\includegraphics{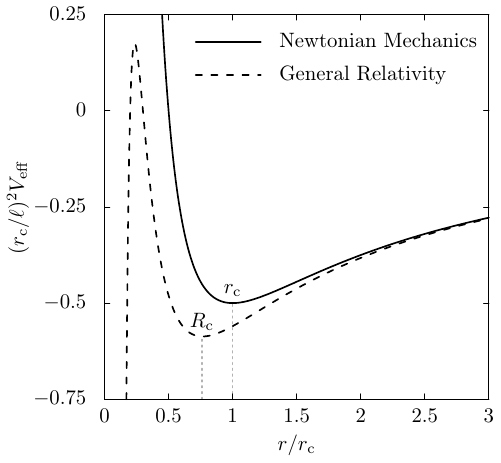}
\caption{\label{fig2} The effective potential commonly defined in the Newtonian limit to General Relativity (dashed) Eq.\,(\ref{eq_Schwarz_Eff_Pot}) is compared to that derived from Newtonian mechanics (solid) with the same angular momentum. The vertical (dotted) lines identify the radii of circular orbits, $R_\text{c}$ and $r_\text{c}$, as calculated using General Relativity and Newtonian mechanics, respectively.  General Relativity predicts a smaller radius of circular orbit, when compared to that predicted by Newtonian mechanics. This reduction in radius of a circular orbit Eq.\,(\ref{eq_S_rc_eff_pot}), $R_\text{c} - r_\text{c} \approx -3\epsilon r_\text{c}$, is three times that predicted by the present treatment using only Special Relativity Eq.\,(\ref{eq_S_coeff_r_mass}), $\bar{r}_\text{c} - r_\text{c} \approx - \epsilon r_\text{c}$.  A curve representing an effective potential including small corrections predicted by Special Relativity is expected to be nearly identical to that for Newtonian mechanics (solid), with a slightly smaller radius of circular orbit, $\bar{r}_\text{c}$.  The value $\epsilon=0.06$ is chosen for purposes of illustration.  Reduction in radius of circular orbit is present for smaller (non-zero) reasonably chosen values of $\epsilon$ as well.}
\end{figure}

%%%%%%%%%%%%%%%%%%%

The general-relativistic (GR) treatment of this problem results in a prediction of 43.0\,arcsec/century \hbox{\cite{LemMon,frisch,carroll,PleKra,MTW,weinberg,rindler,OR,dinverno,HobEfsLas,doggett,stump,ovanesyan,nobili,magnan,deliseo,brillgoel,dean,ashby,wald,hartle,schild,LarCab},} and agrees with the observed precession of perihelia of the inner planets \hbox{\cite{PleKra,MTW,weinberg,rindler,OR,dinverno,HobEfsLas,doggett,stump,ovanesyan,hartle2,stewart,BroCle,sigismondi}}.  Historically, this contribution to the precession of perihelion of Mercury's orbit precisely accounted for the observed discrepancy, serving as the first triumph of the general theory of relativity \cite{einstein0,einstein1,einstein2,einstein3}.  The present approach, using only Special Relativity, accounts for approximately one-third of the observed discrepancy Eq.\,(\ref{eq_prec_num}).

%%%%%%%%%%%%%%%%%%%%

\begin{figure}%[H]
\centering\includegraphics{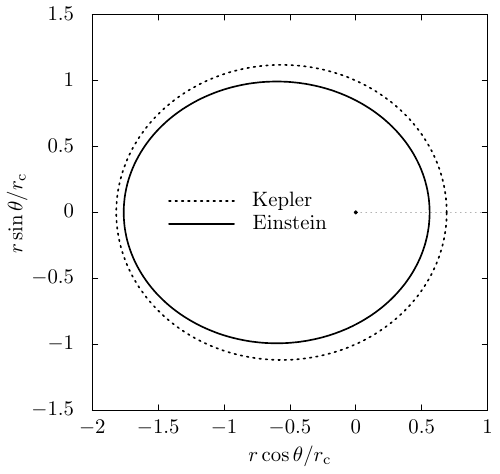}
\caption{\label{fig_rad_pre} A relativistic orbit in a Keplerian limit (solid) Eq.\,(\ref{eq_class_rel_mass2}) is compared to a Keplerian orbit (dotted) Eq.\,(\ref{eq_Newton2}) with the same angular momentum.  Precession of perihelion has been removed from the relativistic orbit equation, Eq.\,(\ref{eq_class_rel_mass2}) with $(1 - \epsilon)\theta \rightarrow \theta$, to emphasize two other characteristics of relativistic orbits---reduced orbital radii and increased eccentricity.  These two characteristics are also predicted by General Relativity Eq.\,(\ref{eq_gen_rel}) in greater magnitude.  These characteristics of relativistic orbits are exaggerated by both the choice of eccentricity $(e=0.45)$ and the relativistic correction parameter $(\epsilon=0.15)$ for purposes of illustration.  Reduced orbital radii and increased eccentricity are present for smaller (non-zero) reasonably chosen values of $e$ and $\epsilon$ as well.  (The same value of $e$ is chosen for both orbits.)}
\end{figure}

The approximate relativistic orbit equation, Eq.\,(\ref{eq_class_rel_mass2}) with $e = 0$, predicts a reduced radius of circular orbit---when compared to that of Newtonian mechanics, Eq.\,(\ref{eq_Newton2}) with $e = 0$.  This characteristic is not discussed in the standard approach to incorporating Special Relativity into the Kepler problem, but is consistent with the GR description.  An effective potential naturally arises in the GR treatment of the central-mass problem \cite{LemMon,rindler,OR,hartle,carroll,HobEfsLas,wald,ovanesyan,MTW2,wang},
\begin{equation}
V_\text{eff} \equiv - \frac{GM}{r} + \frac{\ell^2}{2 r^2} - \frac{GM\ell^2 }{c^2 r^3}, \label{eq_Schwarz_Eff_Pot}
\end{equation}
that reduces to the Newtonian effective potential in the limit $c \rightarrow \infty$.  In the Keplerian limit, the GR angular momentum per unit mass $\ell$ is also taken to be approximately equal to that for a Keplerian orbit \cite{LemMon,MTW,weinberg,rindler,PleKra,OR,dinverno,hartle,carroll,HobEfsLas,ovanesyan,wang,harvey}.  Minimizing $V_\text{eff}$ with respect to $r$ results in the radius of a stable circular orbit,
\begin{equation}
R_\text{c} = \frac{1}{2} r_\text{c} + \frac{1}{2} r_\text{c} \sqrt{1 - 12 \epsilon} \approx r_\text{c} (1 - 3\epsilon), \label{eq_S_rc_eff_pot}
\end{equation}
so that the radius of circular orbit is predicted to be reduced, $R_\text{c} - r_\text{c} \approx -3\epsilon r_\text{c}.$  (There is also an unstable circular orbit, as illustrated in Fig.~\ref{fig2}.)  This reduction in radius of a circular orbit is three times that predicted by the present treatment using only Special Relativity Eq.\,(\ref{eq_S_coeff_r_mass}), for which $\bar{r}_\text{c} - r_\text{c} \approx - \epsilon r_\text{c}$.  Reduced size of an orbit, as predicted by Special Relativity, is illustrated in Fig.~\ref{fig_rad_pre}.

%%%%%%%%%%%%%%%%%%

Many discussions of the GR effective potential Eq.\,(\ref{eq_Schwarz_Eff_Pot}) emphasize relativistic capture.  The $1/r^3$ term in Eq.\,(\ref{eq_Schwarz_Eff_Pot}) contributes negatively to the effective potential, resulting in a finite---rather than infinite---centrifugal barrier and affecting orbits very near the central mass (large-velocity orbits), as illustrated in Fig.~\ref{fig2}.  This purely GR effect is not expected to be described by the approximate orbit equation Eq.\,(\ref{eq_class_rel_mass2}), which is derived using only Special Relativity and implicitly assumes orbits very far from the central mass (small-velocity orbits).

%%%%%%%%%%%%%%%%%%

An additional characteristic of relativistic orbits is that of increased eccentricity.  The relativistic orbit equation Eq.\,(\ref{eq_class_rel_mass2}) predicts increased eccentricity, when compared to a Keplerian orbit Eq.\,(\ref{eq_Newton2}) with the same angular momentum Eq.\,(\ref{eq_S_coeff_e_mass}),  $\bar{e} - e \approx \epsilon e$.  This characteristic of relativistic orbits is not discussed in the standard approach to incorporating Special Relativity into the Kepler problem, but is consistent with the GR description.  The GR orbit equation in this Keplerian limit Eq.\,(\ref{eq_gen_rel}) predicts an increase in eccentricity $\bar{e} - e \approx 3 \epsilon e$, which is three times that predicted by the present treatment using only Special Relativity.  Increased eccentricity of an orbit, as predicted by Special Relativity, is illustrated in Fig.~\ref{fig_rad_pre}.

%%%%%%%%%%%%%%%%%%%%%%%%%
\section{Discussion} \label{sec_discussion}
%%%%%%%%%%%%%%%%%%%%%%%%%

The debate concerning the usage of \textsl{relativistic inertial mass} and \textsl{relativistic gravitational mass} \cite{okun,hecht,manor,khrapco,brown,oas1,oas2,adler,vasiliev} is irrelevant in the present context due to the fundamental incompatibility of Special Relativity and gravitation \cite{MTW3}.  Accordingly, language describing the replacement $m \rightarrow \gamma m$ is intentionally omitted.  Instructors may feel more comfortable using the equivalent replacements $v \rightarrow \gamma v$ to introduce relativistic momentum, and $G \rightarrow \gamma G$ to introduce relativistic gravitational force.  Rather, Lagrangians are constructed using familiar elements of Special Relativity for the purpose of simulating general-relativistic orbital effects using only Special Relativity in a well-defined Keplerian limit.  One purpose of this presentation is to provide students with interesting and tractable problems that arise from small special-relativistic modifications to a familiar problem---Kepler's orbits, the solutions of which provide a qualitative understanding of corrections to Kepler's orbits due to General Relativity.  In this Keplerian limit, these models are supposed to be physical based on the likeness of the equations of motion to those derived using General Relativity \cite{LemMon}.  Another purpose is to present methods by which similar models may be constructed and solved.

%%%%%%%%%%%%%%%%%%%%%%%%%
\subsection{Domain of Validity} \label{sec_domain}
%%%%%%%%%%%%%%%%%%%%%%%%%

The following discussion concerning the derivation, validity, and scope of the approximate special-relativistic orbit equations follows the presentation in Secs.\,\ref{sec_lagrange}~and~\ref{Sec_KeplerianLimit}, in which only relativistic kinetic energy is included.  The arguments and conclusions also apply to the presentation in Sec.\,\ref{sec_vector}, in which both relativistic kinetic energy and relativistic gravitational potential energy are included.  The approximate relativistic orbit equation Eq.\,(\ref{eq_class_rel}) provides small corrections to Kepler's orbits Eq.\,(\ref{eq_Newton2}) due to Special Relativity.  A systematic verification may be carried out by substituting Eq.\,(\ref{eq_class_rel}) into Eq.\,(\ref{eq_SR9}), and only keeping terms of orders $e$, $\epsilon$, and $e\epsilon$. The domain of validity is expressed by subjecting the solution Eq.\,(\ref{eq_class_rel}) to the condition
\begin{equation}
\frac{r_\text{c}}{r} - 1 \ll 1 \label{eq_DOV}
\end{equation}
for the smallest value of $r$. Evaluating the orbit equation Eq.\,(\ref{eq_class_rel_mass2}) at perihelion $r_\text{p}$ results in
\begin{equation}
\frac{r_\text{c}}{r_\text{p}} = \frac{1 + e (1 + \tfrac{1}{2}\epsilon)}{1 - \tfrac{1}{2}\epsilon}.
\end{equation}
Substituting this into Eq.\,(\ref{eq_DOV}) results in the domain of validity
\begin{equation}
e(1 + \tfrac{1}{2}\epsilon) + \epsilon \ll 1. \label{eq_S_1st_valid}
\end{equation}
Therefore, the relativistic eccentricity $\bar{e} = e(1 + \tfrac{1}{2}\epsilon)\ll 1$, and Eq.\,(\ref{eq_class_rel}) is limited to describing relativistic corrections to near-circular (Keplerian) orbits. Also, the relativistic correction $\epsilon\ll 1$, and thus the orbit equation Eq.\,(\ref{eq_class_rel}) is valid only for small relativistic corrections.

%%%%%%%%%%%%%%%%%%%%%%%%%

The correction to Keplerian orbits due to Special Relativity $\lambda \equiv \gamma - 1$ [defined after Eq.\,(\ref{eq_SR_rel_mass}) and in Eq.\,(\ref{eq_approx_lambda2})] is approximated using the first-order series $\gamma \approx 1 + \tfrac{1}{2}(v/c)^2$, and neglecting the radial component of the velocity $v \approx r\dot{\theta}$.  Neglecting the radial component of velocity in the relativistic correction $\lambda$ is consistent with the assumption of approximately Keplerian (near-circular) orbits, and is supported by the condition $r_\text{c}/r - 1 \ll 1$ preceding Eq.\,(\ref{eq_linearized2}).  It is emphasized that the radial component of velocity is neglected only in the relativistic correction $\lambda$; it is not neglected in the derivation of the relativistic equation of motion Eq.\,(\ref{eq_SR_rel_mass}).  That there is no explicit appearance of $\dot{r}$ in the relativistic equation of motion, other than in the definition of $\gamma$, is due to a fortunate cancellation after Eq.\,(\ref{eq_part2}).

%%%%%%%%%%%%%%%%%%%%%%%%%

The presentation in Sec.\,\ref{sec_vector} includes relativistic gravitational potential energy using the replacement $m \rightarrow \gamma m$ in the Newtonian gravitational force.  Although the use of relativistic gravitational mass in the special-relativistic Kepler problem is discouraged by some authors \cite{oas1,*oas2}, there are several useful results in the Keplerian limit.  This relativistic gravitational force is approximated, resulting in a conservative force, from which an approximate relativistic potential energy is derived---thereby enabling the use of the Lagrangian formalism.  The resulting approximate relativistic orbit equation Eq.\,(\ref{eq_class_rel_mass2}) is more accurate, when compared to that derived from General Relativity in the same limit Eq.\,(\ref{eq_gen_rel}), than that derived using only relativistic kinetic energy Eq.\,(\ref{eq_class_rel}).  Specifically, this more accurate orbit equation demonstrates that---in this Keplerian limit---the only consequence of neglecting relativistic gravitational potential energy is that corrections due to Special Relativity are decreased by a factor of two.

%%%%%%%%%%%%%%%%%%%%%%%%%
\subsection{Structure of the Models} \label{sec_structure}
%%%%%%%%%%%%%%%%%%%%%%%%%

These models are appealing because they produce equations of motion that are similar to those derived using General Relativity.  Compare the special-relativistic (SR) equation of motion, Eq.\,(\ref{eq_EOM2_mass}) in Sec.\,\ref{sec_lagrange} [using \hbox{$\gamma GM = GM + GM(\gamma - 1)$],} to the general-relativistic (GR) equation of motion, Eq.\,(10) in Ref.\,\onlinecite{LemMon},
\begin{alignat}{3}
\text{(SR)}& \quad \ell^2\frac{\mathrm{d}^2}{\mathrm{d} \theta^2} \frac{1}{r} &&- GM + \frac{\ell^2}{r} - GM ( \gamma - 1 ) &&= 0 \label{eq_EOM2_mass2} \\
\text{(GR)}& \quad \bar{\ell}^2\frac{\mathrm{d}^2}{\mathrm{d} \varphi^2} \frac{1}{r} &&- GM + \frac{\bar{\ell}^2}{r} - GM \Bigl( \frac{3\bar{\ell}^2}{c^2 r^2} \Bigr) &&= 0. \label{eq_EOM2_gr}
\end{alignat}
The terms in parentheses describe corrections to Newtonian orbits due to Special Relativity Eq.\,(\ref{eq_EOM2_mass2}) and General Relativity Eq.\,(\ref{eq_EOM2_gr});  these terms are zero in the Newtonian limit $c \rightarrow \infty$.  Note that both of these equations are exact in the sense that they are derived directly from Lagrangians without making any approximations.  The SR equation of motion Eq.\,(\ref{eq_EOM2_mass2}) has a correction to Newtonian orbits $GM ( \gamma - 1 )$.  Using the first-order expansion $\gamma \approx 1 + \tfrac{1}{2}(v/c)^2$, neglecting the radial component of velocity $v \approx r \dot{\theta}$, and using angular momentum to eliminate $\dot{\theta}$ results in
\begin{equation}
GM (\gamma - 1) \approx GM \Bigl( \frac{\ell^2}{2 c^2 r^2} \Bigr). \label{eq_corr1}
\end{equation}
Aside from a constant, this term is identical to the term that naturally arises in the GR derivation Eq.\,(\ref{eq_EOM2_gr}), and is directly responsible for the $\tfrac{1}{2} \epsilon$ corrections that appear in the approximate relativistic orbit equation Eq.\,(\ref{eq_class_rel}).  The factor of $\gamma$ that is responsible for this term is a direct result of the $\gamma$ that appears in the definition of angular momentum (per unit mass) Eq.\,(\ref{eq_ang_mom}) \hbox{$\ell \equiv \gamma r^2 \dot{\theta}$.}  Fundamentally, this angular momentum has a $\gamma$ factor due to the definition of relativistic kinetic energy \hbox{$\tilde{T} \equiv -mc^2\gamma^{-1}$.}  The GR angular momentum is found to be simply $\bar{\ell} \equiv r^2 \dot{\varphi}$ [Eq.\,(4) in Ref.\,\onlinecite{LemMon}] because that problem is solved using proper time $\tau$ rather than coordinate time $t$, so that \hbox{$\dot{\varphi} \equiv \rmd \varphi/\rmd \tau$.}  A transformation to coordinate time results in \mbox{$\bar{\ell} \equiv \tildegamma r^2 \dot{\varphi}$,} where \hbox{$\tildegamma \equiv \rmd t/\rmd \tau = [ 1 + 2V(r)/c^2 ]^{-1/2}$,} and \hbox{$V(r) \equiv -GM/r$} is the Newtonian gravitational potential.

The vector calculus formalism in Sec.\,\ref{sec_vector_form} includes a $\gamma$ factor in the definition of the relativistic gravitational force.  The $\gamma$ factor from the relativistic angular momentum propagates through the derivation of the orbit equation as described in the preceding paragraph, resulting in a correction to Newtonian orbits $GM (\gamma^2 - 1)$.  Using the first-order expansion $\gamma^2 \approx 1 + (v/c)^2$, neglecting the radial component of velocity $v \approx r \dot{\theta}$, and using angular momentum to eliminate $\dot{\theta}$ results in
\begin{equation}
GM (\gamma^2 - 1) \approx GM \Bigl( \frac{\ell^2}{c^2 r^2} \Bigr). \label{eq_corr2}
\end{equation}
This is exactly twice the correction term that results from using only relativistic kinetic energy Eq.\,(\ref{eq_corr1}), and is directly responsible for the $\epsilon$ corrections that appear in the approximate relativistic orbit equation, Eqs.\,(\ref{eq_class_rel_mass})~and~(\ref{eq_class_rel_mass2}).  Using the Lagrangian formalism in Sec.\,\ref{sec_lagrange_form}, there is an additional requirement that the relativistic correction to the Newtonian gravitational force does not depend \added{explicitly} on $\theta$ or $\dot{\theta}$, so that the simple $\gamma$ factor appears in the definition of angular momentum.

More generally, any function $f(\gamma)$ may be used in the definition of relativistic gravitational force, provided a first-order expansion of the form $\gamma f(\gamma) - 1 \approx \alpha(v/c)^2$ exists, where $\alpha>0$ is a constant.  For example, choosing a simple power-law dependence \mbox{$f(\gamma) = \gamma^\text{n}$,} so that \hbox{$\tilde{F}_\text{g} \equiv \gamma^\text{n} GMm/r^2$,} results in an approximate relativistic orbit equation identical to Eq.\,(\ref{eq_class_rel}) with the replacement $\epsilon \rightarrow (\text{n}+1) \epsilon$.  Notice the physical condition $\text{n} \ge 0$ that is necessary to insure the proper \textsl{direction} of precession, \textsl{reduced} orbital radii, and \textsl{increased} eccentricity.

Using these types of models, the Keplerian limit is defined precisely by only approximating the correction term that represents a modification of the Keplerian equation of motion.  The approximation of this correction term consists of: a series expansion to first order in $(v/c)^2$;  neglecting the higher-order $\dot{r}^2$ term; and making a change of variable subject to the condition $r_\text{c}/r - 1 \equiv 1/s \ll 1$, allowing the linearization $(r_\text{c}/r)^\text{n} \approx 1 + \text{n}/s$. 

With appropriate conditions and approximations, modifications to Newtonian gravity that depend, more generally, on velocity and radial coordinate may be used.  This is most easily described by example.  A toy model is presented in App.\,\ref{app_math_prob} and discussed in the following section that describes methods---including a more broadly-defined Keplerian limit---that are useful for solving problems with more general modifications to Newtonian gravity.

%%%%%%%%%%%%%%%%%%%%%%%%%
\subsection{A Toy Model} \label{sec_discuss_toy}
%%%%%%%%%%%%%%%%%%%%%%%%%

There are many attempts to motivate a physically appropriate modification to Newtonian gravity in the literature \cite{SinghPatra,bunchaft,hidalgo,phipps3,kurucz,vfg,abci}.  The possibility of simply replacing $m \rightarrow \gamma m$ in the Newtonian gravitational potential energy and using the Lagrangian formalism is explored in App.\,\ref{app_math_prob} and discussed in this section.  This toy model is useful for describing methods for solving problems with more general modifications to Newtonian gravity, for which a more broadly-defined Keplerian limit is needed.  The Lagrangian for this model is
\begin{equation}
L = -mc^2 \gamma^{-1} + \gamma \frac{GMm}{r}.
\end{equation}
Lagrange's equations result in an abstruse set of differential equations, Eqs.\,(\ref{eq_deriv_ang_mom4})~and~(\ref{eq_eom_mass4}), that---using an appropriate approximation---reduce to the equations of motion derived in Sec.\,\ref{sec_vector}, Eqs.\,(\ref{eq_ang_mom2})~and~(\ref{eq_eom2}), thereby reproducing the orbit equation, Eqs.\,(\ref{eq_class_rel_mass})~and~(\ref{eq_class_rel_mass2})
\begin{equation}
\frac{r_\text{c}(1 - \epsilon)}{r} \approx 1 + e(1 + \epsilon)\cos{(1 - \epsilon)\theta}.
\end{equation}

%%%%%%%%%%%%%%%%%%%%%%

Another solution to the differential equations, Eqs.\,(\ref{eq_deriv_ang_mom4})~and~(\ref{eq_eom_mass4}), is derived using a more broadly-defined Keplerian limit.  For near-circular approximately Newtonian orbits, $\dot{r}$ and $\gammadot$ are very slowly varying functions, so that the higher-order terms $\gammadot \dot{r}$ and $(\gamma/c)^2 V(r) \gamma \dot{r}^2/r$ are taken to be negligible.  The result is the Newtonian equation of motion with a relativistic correction term Eq.\,(\ref{eq_orbit_mass})
\begin{equation}
\tilde{\ell}^2\frac{\mathrm{d}^2}{\mathrm{d} \theta^2} \frac{1}{r} - GM + \frac{\tilde{\ell}^2}{r} - GM \tilde{\lambda} = 0,
\end{equation}
where
\begin{equation}
\tilde{\lambda} \equiv \gamma^2 \left[ 1 - (\gamma/c)^2 V(r) \right] - 1,
\end{equation}
$\tilde{\ell} \equiv \gamma r^2 \dot{\theta} \left[ 1 - (\gamma/c)^2 V(r) \right]$ is a constant of motion, and $V(r) \equiv -GM/r$ is the Newtonian gravitational potential.  Compare this equation of motion to Eqs.\,(\ref{eq_EOM2_mass2})~and~(\ref{eq_EOM2_gr}).  This equation of motion has a small relativistic correction to Keplerian orbits Eq.\,(\ref{eq_toy_lambda})
\begin{align}
GM \tilde{\lambda} \approx GM \biggl[ \frac{\tilde{\ell}^2}{c^2 r^2} - \frac{V(r)}{c^2} \biggr].
\end{align}
Compare this to the correction terms used in the two simple models, Eqs.\,(\ref{eq_corr1})~and~(\ref{eq_corr2}).  The corresponding orbit equation in the Keplerian limit Eq.\,(\ref{eq_SSR3}),
\begin{gather}
\frac{\tilde{r}_\text{c}(1-2\tilde{\epsilon})}{r} \approx 1 + e(1 + \tilde{\epsilon})\cos{(1 - \tfrac{3}{2} \tilde{\epsilon})\theta}, \label{eq_SSR4}
\end{gather}
does not have the symmetry of those derived using the two simple models, Eqs.\,(\ref{eq_class_rel})~and~(\ref{eq_class_rel_mass2}).  The methods and approximations used to derive this orbit equation may be applied to other, more physical, models that include relativistic corrections to Kepler's orbits.

%%%%%%%%%%%%%%%%%%%%%%%%%
\section{Conclusion}
%%%%%%%%%%%%%%%%%%%%%%%%%

The Lagrangian formalism is useful for describing small relativistic corrections to Kepler's orbits.  The simplest model includes only relativistic \replaced{kinetic energy}{inertial mass} using the replacement $m \rightarrow \gamma m$ in the Newtonian linear momentum, $p = \gamma m v$.  A solution to the corresponding equations of motion in a Keplerian limit results in an approximate relativistic orbit equation Eq.\,(\ref{eq_class_rel}) that has the same form as that derived from General Relativity in this limit Eq.\,(\ref{eq_gen_rel}) and is easily compared to that describing Kepler's orbits Eq.\,(\ref{eq_Newton2}).  This form is that of elliptical orbits of Newtonian mechanics with corrections to radius and eccentricity, and exhibiting precession.  Specifically, the approximate relativistic orbit equation clearly describes three characteristics of relativistic orbits:  precession of perihelion; reduced radius of circular orbit; and increased eccentricity.  The predicted rate of precession of perihelion of Mercury is in agreement with that of established calculations using only Special Relativity.  Each of these characteristics of relativistic Keplerian orbits is exactly one-sixth of the corresponding correction described by General Relativity in this limit---providing a qualitative description of corrections to Keplerian orbits due to General Relativity.

%%%%%%%%%%%%%%%%%%%%%%%%%

This simple model is improved upon by including relativistic gravitational force using the replacement $m \rightarrow \gamma m$ in Newtonian gravity, $\tilde{F}_\text{g} = \gamma GMm/r^2$.  An approximation consistent with the Keplerian limit results in a conservative force, from which a relativistic potential energy is derived that is useful in a Lagrangian formulation of the special-relativistic Kepler problem.  A solution of the corresponding equations of motion in a Keplerian limit results in an approximate relativistic orbit equation Eq.\,(\ref{eq_class_rel_mass2}) that has the same form as that derived using the simplest model Eq.\,(\ref{eq_class_rel}), thereby describing the same three characteristics of relativistic orbits.  Each of these characteristics of relativistic Keplerian orbits is exactly one-third of the corresponding correction described by General Relativity in this limit Eq.\,(\ref{eq_gen_rel}).

%%%%%%%%%%%%%%%%%%%%%%%%%

The Lagrangian formalism applied to the special-relativistic Kepler problem is instructive, providing several challenges appropriate for an introductory classical mechanics course, including:  solve Newton's force equation using vector calculus to verify the unfamiliar relativistic kinetic energy term in the Lagrangian---as outlined in the last paragraph of Sec.\,\ref{Sec_KeplerianLimit} and in Sec.\,\ref{sec_vector_form};  derive a potential energy function from a conservative force;  apply Lagrange's equations to derive the conserved relativistic angular momentum and equation of motion;  and transform and solve a differential equation to derive an approximate relativistic orbit equation in terms of planar coordinates.  This approach also provides an opportunity to use less familiar problem solving strategies, including:  variable transformations to cast the differential equation into familiar form;  approximation methods that simplify the differential equation;  and usage of the correspondence principle to identify a constant of integration.  Most importantly, students are rewarded with a clear understanding that a small relativistic modification to a familiar problem results in an approximate relativistic orbit equation that clearly demonstrates that relativity is responsible for a small contribution to perihelic precession, and the satisfaction of calculating that contribution.

%%%%%%%%%%%%%%%%%%%%%%%%%

These models are appealing because they are easy to motivate and, most importantly, they produce equations of motion that are similar to those derived using General Relativity, as discussed in Sec.\,\ref{sec_structure}.  A larger class of models may be solved using similar methods and approximations.  This is demonstrated using a toy model in App.\,\ref{app_math_prob} and in Sec.\,\ref{sec_discuss_toy}, wherein a more broadly-defined Keplerian limit is described.  Exact solutions of the special-relativistic Kepler problem require a thorough understanding of special relativistic mechanics \cite{rindler2,synge} and are, therefore, inaccessible to many undergraduate physics majors.  The present approach and method of solution is understandable to nonspecialists, including undergraduate physics majors whom have not had a course dedicated to relativity.

%%%%%%%%%%%%%%%%%%%%%%%%%
%\begin{acknowledgments}
%The authors would like to thank Shane Burns, Katherine Mondragon, and Patricia Purdue for their valuable comments, suggestions, and corrections.
%\end{acknowledgments}
%%%%%%%%%%%%%%%%%%%%%%%%%

\renewcommand{\baselinestretch}{1.02}

\renewcommand{\baselinestretch}{1.08}\small\normalsize
%%%%%%%%%%%%%%%%%%%%%%%%%
{\appendix\section{A Toy Model}\label{app_math_prob}
%%%%%%%%%%%%%%%%%%%%%%%%%

The possibility of including relativistic gravitational potential by simply replacing $m \rightarrow \gamma m$ in the Newtonian gravitational potential energy and using the Lagrangian formalism is explored.  This toy model is useful for describing methods for solving problems with more general modifications to Newtonian gravity, for which a more broadly-defined Keplerian limit is needed.  The Lagrangian for this model is
\begin{equation}
L = -mc^2 \gamma^{-1} + \gamma \frac{GMm}{r}.
\end{equation}
Lagrange's equations result in an abstruse set of differential equations
\begin{gather}
\dfrac{\rmd}{\rmd t} \left\{ \gamma r^2 \dot{\theta} \left[ 1 - (\gamma/c)^2 V(r) \right] \right\} = 0, \label{eq_deriv_ang_mom4} \\
\text{and} \nonumber \\
\begin{multlined}[0.75\columnwidth]
\gamma \ddot{r} \left[ 1 - (\gamma/c)^2 V(r) \right] + \gammadot \dot{r} \left[ 1 - 3(\gamma/c)^2 V(r) \right] \\[1ex]
+ \gamma \frac{GM}{r^2} - \gamma r \dot{\theta}^2 \left[ 1 - (\gamma/c)^2 V(r) \right] \\
+ (\gamma/c)^2 V(r) \gamma \frac{\dot{r}^2}{r} =0,
\end{multlined} \label{eq_eom_mass4}
\end{gather}
where $V(r) \equiv -GM/r$ is the Newtonian gravitational potential.  Mercury's orbit is approximately circular with radius of the same order of magnitude as its semimajor axis $r \sim a \approx 5.79 \times 10^{10}\,\text{m}$, and its velocity is very small when compared to the speed of light, so that $(\gamma/c)^2 V(r) \sim V(a)/c^2 \sim -10^{-8}$.  Ignoring terms proportional to $(\gamma/c)^2 V(r)$ results in the equations of motion derived in Sec.\,\ref{sec_vector}, Eqs.\,(\ref{eq_ang_mom2})~and~(\ref{eq_eom2}), thereby reproducing the orbit equation, Eqs.\,(\ref{eq_class_rel_mass})~and~(\ref{eq_class_rel_mass2})
\begin{equation}
\frac{r_\text{c}(1 - \epsilon)}{r} \approx 1 + e(1 + \epsilon)\cos{(1 - \epsilon)\theta}.
\end{equation}

%%%%%%%%%%%%%%%%%%%

Another solution to the equations of motion, Eqs.\,(\ref{eq_deriv_ang_mom4})~and~(\ref{eq_eom_mass4}), is derived using a more broadly-defined Keplerian limit.  For near-circular approximately Newtonian orbits, $\dot{r}$ and $\gammadot$ are very slowly varying functions, so that the higher-order terms $\gammadot \dot{r}$ and $(\gamma/c)^2 V(r) \gamma \dot{r}^2/r$ are taken to be negligible.  [For Mercury's orbit $(\gamma/c)^2 V(a) \gamma/a \sim -10^{-18}\,\text{m}^{-1}$.]
\begin{gather}
\tilde{\ell} \equiv \gamma r^2 \dot{\theta} \left[ 1 - (\gamma/c)^2 V(r) \right] = \text{constant}, \label{eq_ang_mom_mass2} \\
\text{and} \nonumber \\
\begin{multlined}[0.75\columnwidth]
\gamma \ddot{r} \left[ 1 - (\gamma/c)^2 V(r) \right] + \gamma \frac{GM}{r^2} \\[0.5ex]
- \gamma r \dot{\theta}^2 \left[ 1 - (\gamma/c)^2 V(r) \right] \approx 0.
\end{multlined} \label{eq_eom_mass2}
\end{gather}
Conservation of angular momentum Eq.\,(\ref{eq_ang_mom_mass2}) is used to eliminate the explicit occurrence of $\dot{\theta}$ in the equation of motion Eq.\,(\ref{eq_eom_mass2})
\begin{equation}
\gamma r \dot{\theta}^2 \left[ 1 - (\gamma/c)^2 V(r) \right] = \frac{\tilde{\ell}^2}{\gamma r^3 \left[ 1 - (\gamma/c)^2 V(r) \right]}. \label{eq_part1_mass2}
\end{equation}
Time is eliminated by successive applications of the chain rule, together with the conserved angular momentum;
\begin{equation}
\dot{r} = - \frac{\tilde{\ell}}{\gamma [ 1 - (\gamma/c)^2 V(r) ]} \frac{\rmd}{\rmd \theta} \frac{1}{r},
\end{equation}
and, therefore, (again taking $\gammadot \dot{r}$ to be negligible)
\begin{equation}
\gamma \ddot{r} [ 1 - (\gamma/c)^2 V(r) ] \approx - \frac{\tilde{\ell}^2}{\gamma [ 1 - (\gamma/c)^2 V(r) ] r^2} \frac{\rmd^2}{\rmd \theta^2} \frac{1}{r}. \label{eq_chain}
\end{equation}
Substituting Eqs.\,(\ref{eq_part1_mass2})~and~(\ref{eq_chain}) into the equation of motion Eq.\,(\ref{eq_eom_mass2}) results in
\begin{equation}
\tilde{\ell}^2\frac{\mathrm{d}^2}{\mathrm{d} \theta^2} \frac{1}{r} - \gamma^2 [ 1 - (\gamma/c)^2 V(r) ] GM + \frac{\tilde{\ell}^2}{r} = 0. \label{eq_orbit_mass}
\end{equation}
Anticipate a solution of Eq.\,(\ref{eq_orbit_mass}) that is near Keplerian and introduce the radius of a circular orbit for a nonrelativistic particle with the same angular momentum, $\tilde{r}_\text{c} \equiv \tilde{\ell}^2/GM $.  The result is
\begin{equation}
\frac{\mathrm{d}^2}{\mathrm{d} \theta^2} \frac{\tilde{r}_\text{c}}{r} + \frac{\tilde{r}_\text{c}}{r} = 1 + \tilde{\lambda}, \label{eq_SR_mass}
\end{equation}
where $\tilde{\lambda} \equiv \gamma^2 [ 1 - (\gamma/c)^2 V(r) ]-1$ is a correction to Newtonian orbits due to Special Relativity.  [Tilde notation is used to emphasize that quantities depend on the exact conserved quantity $\tilde{\ell}$ defined in Eq.\,(\ref{eq_ang_mom_mass2}), rather than the angular momentum $\ell$ defined in Eq.\,(\ref{eq_ang_mom2}).]

%%%%%%%%%%%%%%%%%%%

The orbit equation is derived following the method described in Sec.\,\ref{Sec_KeplerianLimit}.  The correction term $\tilde{\lambda}$ is approximated by expanding to first-order in $1/c^2$, neglecting the radial component of velocity, and using angular momentum to eliminate $\dot{\theta}$
\begin{equation}
\tilde{\lambda} \approx ( \tilde{\ell}/rc )^2 - V(r)/c^2. \label{eq_toy_lambda}
\end{equation}
The equation of motion Eq.\,(\ref{eq_SR_mass}) is now expressed approximately as
\begin{equation}
\frac{\mathrm{d}^2}{\mathrm{d} \theta^2} \frac{\tilde{r}_\text{c}}{r} + \frac{\tilde{r}_\text{c}}{r} \approx 1 + \tilde{\epsilon} \dfrac{\tilde{r}_\text{c}}{r} + \tilde{\epsilon} \Bigl( \dfrac{\tilde{r}_\text{c}}{r} \Bigr)^{\!2}, \label{eq_SR7}
\end{equation}
where $\tilde{\epsilon} \equiv (GM/\tilde{\ell}c)^2$.  An orbit equation is derived, as described in Sec.\,\ref{Sec_KeplerianLimit}.  The equation of motion Eq.\,(\ref{eq_SR7}) is linearized using $r_\text{c}/r = 1 + 1/s$, and $(r_\text{c}/r)^2 \approx 1 + 2/s$.  The additional change of variable $\alpha \equiv \theta\sqrt{1-3\tilde{\epsilon}}$ results in the familiar differential equation
\begin{equation}
\frac{\mathrm{d}^2}{\mathrm{d} {\alpha}^2} \frac{s_\text{c}}{s} + \frac{s_\text{c}}{s} \approx 1,
\end{equation}%
where $s_\text{c} \equiv (1-3\tilde{\epsilon})/(2\tilde{\epsilon})$.  The solution is similar to that of Eq.\,(\ref{eq_Newton1})
\begin{equation}
\frac{s_\text{c}}{s} \approx 1   + A\cos{\alpha},
\end{equation}
where $A$ is an arbitrary constant of integration.  In terms of the original coordinates, and defining $e \equiv 2\epsilon A$, an orbit equation in the Keplerian limit is described concisely by
\begin{gather}
\frac{\tilde{r}_\text{c}(1-2\tilde{\epsilon})}{r} \approx 1 + e(1 + \tilde{\epsilon})\cos{(1 - \tfrac{3}{2} \tilde{\epsilon})\theta}.\label{eq_SSR3}
\end{gather}
This orbit equation does not have the symmetry of those derived using the two simple models, Eqs.\,(\ref{eq_class_rel})~and~(\ref{eq_class_rel_mass2}).  The methods and approximations used to derive this orbit equation may be applied to other, more physical, models that include relativistic corrections to Kepler's orbits.  See the discussions in Sec.\,\ref{sec_structure} and Sec.\,\ref{sec_discuss_toy}.
}


\small\normalsize
\begin{thebibliography}{99}

\bibitem{einstein0} A.~Einstein, ``Erkl\"{a}rung der Perihelbewegung des Merkur aus der allgemeinen Relativit\"{a}tstheorie,'' Sitzungsber. Preuss. Akad. Wiss. Berlin (Math. Phys.) \textbf{1915}, 831--839 (1915).

\bibitem{einstein1} A.~Einstein, ``Explanation of the perihelion motion of mercury from the general theory of relativity,'' in \textit{The Collected Papers of Albert Einstein}, translated by A. Engel (Princeton University Press, Princeton, 1997), Vol. 6, pp.\,112--116. This article is the English translation of Ref.\,\onlinecite{einstein0}.

\bibitem{einstein2} A.~Einstein, ``Die Grundlage der allgemeinen Relativit\"{a}tstheorie; von A. Einstein,'' Ann. d. Phys. \textbf{354} (7), 769--822 (1916).

\bibitem{einstein3} Reference~\onlinecite{einstein1}, pp.\,146--200. This article is the English translation of Ref.~\onlinecite{einstein2}.

\bibitem{schwarzschild} K.~Schwarzschild, ``\"{U}ber das Gravitationsfeld eines Massenpunktes nach der Einsteinschen Theorie,'' Sitzungsber. Preuss. Akad. Wiss., Phys.-Math. Kl. \textbf{1916}, 189--196 (1916).  Reprinted in translation as ``On the Gravitational Field of a Mass Point according to Einstein's Theory,'' \href{http://arxiv.org/pdf/physics/9905030v1.pdf}{arXiv:\,physics/9905030v1 [physics.hist-ph]}.

\bibitem{droste} J.~Droste, ``Het veld van een enkel centrum in Einstein's theorie der zwaartekracht, en de beweging van een stoffelijk punt in dat veld,'' Versl. Akad. Amst. \textbf{25}, 163--180 (1916-1917). Reprinted in translation as ``The field of a single centre in Einstein's theory of gravitation, and the motion of a particle in that field,'' Proc. K. Ned. Akad. Wetensch. \textbf{19} (1), 197--215 (1917), \url{<adsabs.harvard.edu/abs/1917KNAB...19..197D>}.

\bibitem{goldstein2} H.~Goldstein, C.~Poole, and J.~Safko, \textit{Classical Mechanics} (Addison Wesley, San Francisco, 2002), 3rd ed., pp.\,312--314 and p.\,332 (Exercise 26).

\bibitem{jose} J.\,V.~Jos\'{e} and E.\,J.~Saletan, \textit{Classical Dynamics: A Contemporary Approach} (Cambridge University Press, Cambridge, 1998), pp.\,209--212 and pp.\,276--277 \hbox{(Problem\,11).}

\bibitem{peters}  P.\,C.~Peters, ``Comment on `Mercury's precession according to special relativity' [Am. J. Phys. \textbf{54}, 245 (1986)],'' {Am. J. Phys.} \textbf{55} (8), 757 (1987).

\bibitem{phipps1}  T.\,Phipps, Jr., ``Response to `Comment on ``Mercury's precession according to special relativity,'''[Am. J. Phys. \textbf{55}, 757 (1987)],'' {Am. J. Phys.} \textbf{55} (8), 758 (1987).

\bibitem{phipps2}  T.~Phipps, Jr., ``Mercury's precession according to special relativity,'' {Am. J. Phys.} \textbf{54} (3), 245 (1986).

%\newpage
\bibitem{jia} L.~Jia, ``Approximate Kepler's Elliptic Orbits with the Relativistic Effects,'' Int. J. Astron. Astrophys. \textbf{3}, 29--33 (2013).

\bibitem{goldstein} Reference~\onlinecite{goldstein2}, pp.\,536--539.

\bibitem{TM1} S.\,T.~Thornton and J.\,B. Marion, \textit{Classical Dynamics of Particles and Systems} (Thomson Brooks/Cole, Belmont, CA, 2004), pp.\,292--293 and pp.\,312--316.

\bibitem{barger} V.\,D.~Barger and M.\,G.~Olsson, \textit{Classical Mechanics: A Modern Perspective} (McGraw-Hill, Inc., New York, 1995), pp.\,306--309.

\bibitem{fowles} G.\,R.~Fowles, \textit{Analytical Mechanics} (Holt, Rinehart and Winston, New York, 1977), pp.\,161--164.

\bibitem{hand} L.\,N.~Hand and J.\,D.~Finch, \textit{Analytical Mechanics} (Cambridge University Press, Cambridge, 1998), pp.\,420--422 (Problems\,18~and~19).

\bibitem{TM2} Reference~\onlinecite{TM1}, pp.\,578--579.

\bibitem{barger2} Reference~\onlinecite{barger}, p.\,366 (Problem\,10-12).

\bibitem{potgieter} J.\,M.~Potgieter, ``Derivation of the equations of Lagrange for a relativistic particle,'' Am. J. Phys. \textbf{51} (1), 77 (1983).

\bibitem{DesEri} E.\,A.~Desloge and E.~Eriksen, ``Lagrange's equations of motion for a relativistic particle,'' Am. J. Phys. \textbf{53} (1), 83--84 (1985).

\bibitem{HuangLin} Y.-S.~Huang and C.-L.~Lin, ``A systematic method to determine the Lagrangian directly from the equations of motion,'' Am. J. Phys. \textbf{70} (7), 741--743 (2002).

\bibitem{SonMas} S.~Sonego and M.~Pin, ``Deriving relativistic momentum and energy,'' Eur. J. Phys. \textbf{26}, 33--45 (2005).

\bibitem{LemMon2} T.\,J.~Lemmon and A.\,R.~Mondragon, ``First-Order Special Relativistic Corrections to Kepler's Orbits'' (2010) unpublished, \href{http://arxiv.org/pdf/1012.5438v1.pdf}{arXiv:\,1012.5438v1 [astro-ph.EP]}.

\bibitem{goldstein4} Reference~\onlinecite{goldstein2}, p.\,87.

\bibitem{jose2} Reference~\onlinecite{jose}, p.\,84.

\bibitem{TM4} Reference~\onlinecite{TM1}, p.\,292.

\bibitem{fowles2} Reference~\onlinecite{fowles}, p.\,148.

\bibitem{hamill} P.~Hamill, \textit{Intermediate Dynamics} (Jones and Bartlett Publishers, LLC, Sudbury, MA, 2010), p.\,330.

\bibitem{goldstein3} Reference~\onlinecite{goldstein2}, pp.\,92--96.

\bibitem{jose3} Reference~\onlinecite{jose}, pp.\,84--86.

\bibitem{TM3} Reference~\onlinecite{TM1}, pp.\,300--304.

\bibitem{fowles3} Reference~\onlinecite{fowles}, pp.\,137--157.

\bibitem{hamill2} Reference~\onlinecite{hamill}, pp.\,307--343.

\bibitem{smiths} P.~Smith and R.\,C.~Smith, \textit{Mechanics} (John Wiley \& Sons, Chichester, 1990), pp.\,195--202.

\newpage
\bibitem{LemMon} T.\,J.~Lemmon and A.\,R.~Mondragon, ``Alternative derivation of the relativistic contribution to perihelic precession,'' Am. J. Phys. \textbf{77} (10), 890--893 (2009); \href{http://arxiv.org/pdf/0906.1221v2.pdf}{arXiv:\,0906.1221v2 [astro-ph.EP]}.

\bibitem{SinghPatra} A.~Singh and B.\,K.~Patra, ``Relativistic corrections to the central force problem in a generalized potential approach,'' (2014), \href{http://arxiv.org/pdf/1404.2940v2.pdf}{arXiv:\,1404.2940v2 [physics.class-ph]}.

\bibitem{bunchaft} F.~Bunchaft and S.~Carneiro, ``Weber-like interactions and energy conservation,'' (1997), \href{http://arxiv.org/pdf/gr-qc/9708047v1.pdf}{arXiv:\,gr-qc/9708047v1}.

\bibitem{hidalgo} R.\,A.\,V.~Hidalgo-Gato, ``Towards an extension of 1905 relativistic dynamics with a variable rest mass measuring potential energy,'' (2012), \href{https://arxiv.org/ftp/arxiv/papers/1210/1210.4157.pdf}{arXiv:\,1210.4157v1 [physics.gen-ph]}.

\bibitem{phipps3} T.\,E.~Phipps, Jr., ``On Gerber's Velocity-dependent Gravitational Potential,'' (2004) unpublished, \url{http://studylib.net/doc/5885051/on-gerber-s-velocity-dependent-gravitational-potential}.

\bibitem{kurucz} R.\,L.~Kurucz, ``The Precession of Mercury and the Deflection of Starlight from Special Relativity Alone,'' (2009) unpublished, \url{kurucz.harvard.edu/papers/deflection/deflection.pdf}.

\bibitem{vfg} C.\,G.~Vayenas, A.~Fokas, and D.~Grigoriou, ``Gravitational mass and Newton's universal gravitational law under relativistic conditions,'' J. Phys: Conf. Series \textbf{633} (2015) 012033.

\bibitem{abci} M.\,A.~Abramowicz, A.\,M.~Beloborodov, X.~Chen, and I.\,V.~Igumenshchev, ``Special Relativity and pseudo-Newtonian potential,'' (1996) unpublished, \href{http://arxiv.org/pdf/astro-ph/9601115v1.pdf}{arXiv:\,astro-ph/9601115v1}.

\bibitem{frisch} D.\,H.~Frisch, ``Simple aspects of post-Newtonian gravitation,'' Am. J. Phys. \textbf{58} (4), 332--337 (1990).

\bibitem{carroll} S.\,M.~Carroll, \textit{An Introduction to General Relativity: Spacetime and Geometry} (Addison-Wesley, San Francisco, 2004), pp.\,205--216.

\bibitem{PleKra} J.~Pleba\'{n}ski and A.~Krasi\'{n}ski, \textit{An Introduction to General Relativity and Cosmology} (Cambridge University Press, Cambridge, 2006), pp.\,176--182.

\bibitem{MTW} C.\,W.~Misner, K.\,S.~Thorne, and J.\,A.~Wheeler, \textit{Gravitation} (Freeman, San Francisco, 1973), pp.\,1110--1115.

\bibitem{weinberg} S.~Weinberg, \textit{Gravitation and Cosmology} (John Wiley \& Sons, New York, 1972), pp.\,194--201.

\bibitem{rindler} W.~Rindler, \textit{Relativity: Special, General, and Cosmological} (Oxford University Press, New York, 2001), pp.\,238--245.

\bibitem{OR} H.~Ohanian and R.~Ruffini, \textit{Gravitation and Spacetime} (W. W. Norton \& Company, New York, 1994), pp.\,401--408.

\bibitem{dinverno} R.~D'Inverno, \textit{Introducing Einstein's Relativity} (Oxford University Press, New York, 1995), pp.\,195--198.

\bibitem{HobEfsLas} M.\,P.~Hobson, G.\,P.~Efstathiou, and A.\,N.~Lasenby, \textit{General Relativity: An Introduction for Physicists} (Cambridge University Press, Cambridge, 2006), pp.\,205--216 and pp.\,230--233.

\bibitem{doggett}  K.~Doggett, ``Comment on `Precession of the perihelion of Mercury,' by Daniel R. Stump [Am. J. Phys. \textbf{56}, 1097--1098 (1988)],'' {Am. J. Phys.} \textbf{59} (9), 851--852 (1991).

%\newpage
\bibitem{stump} D.\,R.~Stump, ``Precession of the perihelion of Mercury,'' Am. J. Phys. \textbf{56} (12), 1097--1098 (1988).

\bibitem{ovanesyan} G.~Ovanesyan, ``Derivation of relativistic corrections to bounded and unbounded motion in a weak gravitational field by integrating the equations of motion,'' Am. J. Phys. \textbf{71} (9), 912--916 (2003).

\bibitem{nobili} A.\,M.~Nobili and I.\,W.~Roxburgh, ``Simulation of general relativistic corrections in long term numerical integrations of planetary orbits,'' in \textit{Relativity in Celestial Mechanics and Astrometry: High Precision Dynamical Thoeries and Observational Verifications}, J.~Kovalevsky and V.\,A.~Brumberg (IAU, Dordrecht, 1986), pp.\,105--110, \url{<adsabs.harvard.edu/abs/1986IAUS..114..105N>}.

\bibitem{magnan} C.~Magnan, ``Complete calculations of the perihelion precession of Mercury and the deflection of light by the Sun in General Relativity,'' (2007) unpublished, \href{http://arxiv.org/pdf/0712.3709v1.pdf}{arXiv:\,0712.3709v1 [gr-qc]}.

\bibitem{deliseo} M.\,M.~D'Eliseo, ``The first-order orbital equation,'' Am. J. Phys. \textbf{75} (4), 352--355 (2007).

\bibitem{brillgoel} D.\,R.~Brill and D.~Goel, ``Light bending and perihelion precession: A unified approach,'' Am. J. Phys. \textbf{67} (4), 316--319 (1999).

\bibitem{dean} B.~Dean, ``Phase-plane analysis of perihelion precession and Schwarzschild orbital dynamics,'' Am. J. Phys. \textbf{67} (1), 78--86 (1999).

\bibitem{ashby} N.~Ashby, ``Planetary perturbation equations based on relativistic Keplerian motion,'' in \textit{Relativity in Celestial Mechanics and Astrometry: High Precision Dynamical Thoeries and Observational Verifications}, edited by J.~Kovalevsky and V.\,A.~Brumberg (IAU, Dordrecht, 1986), pp.\,41--52, \url{<adsabs.harvard.edu/abs/1986IAUS..114...41A>}.

\bibitem{wald} R.\,M.~Wald, \textit{General Relativity} (University of Chicago Press, Chicago, 1984), pp.\,136--143.

\bibitem{hartle} J.\,B.~Hartle, \textit{Gravity: An Introduction to Einstein's General Relativity} (Addison-Wesley, San Francisco, 2003), pp.\,191--204.

\bibitem{schild} A.~Schild, ``Equivalence principle and red-shift measurements,'' Am. J. Phys. \textbf{28} (9), 778--780 (1960).

\bibitem{LarCab} A.~Larra\~{n}aga and L.~Cabarique, ``Advance of Planetary Perihelion in Post-Newtonian Gravity,'' Bulg. J. Phys. \textbf{30}, 1--7 (2005), \href{http://arxiv.org/pdf/1202.4951v1.pdf}{arXiv:\,1202.4951v1 [gr-qc]}.

\bibitem{hartle2} Reference~\onlinecite{hartle}, pp.\,230--232.

\bibitem{stewart} M.\,G.~Stewart, ``Precession of the perihelion of Mercury's orbit,'' Am. J. Phys. \textbf{73} (8), 730--734 (2005).

\bibitem{BroCle} D.~Brouwer and G.\,M.~Clemence, ``Orbits and masses of planets and satellites,'' in \textit{The Solar System: Planets and Satellites}, edited by G.\,P.~Kuiper and B.\,M.~Middlehurst (University of Chicago Press, Chicago, 1961), Vol. III, pp.\,31--94.

\bibitem{sigismondi} C.~Sigismondi, ``Astrometry and relativity,'' {Nuovo Cim. B} \textbf{120}, 1169--1180 (2005).

\bibitem{MTW2} Reference~\onlinecite{MTW}, pp.\,660--662, Box~25.6.

\bibitem{wang} F.\,Y.-H. Wang, ``Relativistic orbits with computer algebra,'' Am. J. Phys. \textbf{72} (8), 1040--1044 (2004).

\bibitem{harvey}  A.~Harvey, ``Newtonian limit for a geodesic in a Schwarzschild field,'' Am. J. Phys. \textbf{46} (9), 928--929 (1978).

\bibitem{okun} L.~Okun, ``The Concept of Mass,'' Physics Today (June, 1989), 31--36.

\bibitem{hecht} E.~Hecht, ``Einstein on mass and energy,'' Am. J. Phys. \textbf{77} (9), 799--806 (2009).

\bibitem{manor} E.\,P.~Manor, ``Gravity, Not Mass Increases with Velocity,'' J. Mod. Phys. \textbf{6}, 1407--1411 (2015), \url{http://dx.doi.org/10.4236/jmp.2015.610145}.

\bibitem{khrapco} R.\,I.~Khrapco, ``Rest mass or inertial mass?,'' (2001) unpublished, \href{http://arxiv.org/pdf/physics/0103008v2.pdf}{arXiv:\,physics/0103008v2 [physics.gen-ph]}.

\bibitem{brown} P.\,M.~Brown, ``On the concept of mass in relativity,'' (2007) unpublished, \href{https://arxiv.org/ftp/arxiv/papers/0709/0709.0687.pdf}{arXiv:\,0709.0687v2 [physics.gen-ph]}.

\bibitem{oas1} G.~Oas, ``On the abuse and use of relativistic mass,'' (2005), \href{http://arxiv.org/pdf/physics/0504110v2.pdf}{arXiv:\,physics/0504110v2 [physics.ed-ph]}.

\bibitem{oas2} G.~Oas, ``On the Use of Relativistic Mass in Various Published Works,'' (2005), \href{http://arxiv.org/pdf/physics/0504111v1.pdf}{arXiv:\,physics/0504111v1 [physics.ed-ph]}.

\bibitem{adler} C.\,G.~Adler, ``Does mass really depend on velocity, dad?,'' Am. J. Phys. \textbf{55} (8), 739--743 (1987).

\bibitem{vasiliev} S.\,A.~Vasiliev, ``On the Notion of the Measure of Inertia in the Special Relativity Theory,'' App. Phys. Res. \textbf{4} (2), 136--143 (2012).

\bibitem{MTW3} Reference~\onlinecite{MTW}, ch.\,7.

\bibitem{rindler2} W.~Rindler, \textit{Special Relativity} (Oliver and Boyd, London, 1960), pp.\,79--103.

\bibitem{synge} J.\,L.~Synge, \textit{Relativity: The Special Theory} (North-Holland Publishing Company, Amsterdam, 1958), pp.\,394--399.
\end{thebibliography}
\end{document}